\documentclass[10pt,twocolumn]{article}

\usepackage[top=0.75in,bottom=1in,left=0.75in,right=0.75in]{geometry}
\usepackage{times}
\usepackage{amsmath,amssymb}
\usepackage{graphicx}
\usepackage{booktabs}
\usepackage{hyperref}
\usepackage{url}
\usepackage{microtype}
\usepackage{array}
\usepackage{titlesec}
\usepackage{enumitem}
\usepackage{dblfloatfix}

\titlespacing*{\section}{0pt}{8pt}{4pt}
\titlespacing*{\subsection}{0pt}{6pt}{3pt}

\hypersetup{colorlinks=true,linkcolor=blue,citecolor=blue,urlcolor=blue}

\title{\textbf{A Large-Scale Empirical Comparison of Meta-Learners and Causal Forests for Heterogeneous Treatment Effect Estimation in Marketing Uplift Modeling}}

\author{Aman Singh}
\date{}

\begin{document}
\maketitle

\begin{abstract}
Estimating Conditional Average Treatment Effects (CATE) at the individual level is central to precision marketing, yet systematic benchmarking of uplift modeling methods at industrial scale remains limited. We present \textbf{UpliftBench}, an empirical evaluation of four CATE estimators: S-Learner, T-Learner, X-Learner (all with LightGBM base learners), and Causal Forest (EconML), applied to the Criteo Uplift v2.1 dataset comprising 13.98 million customer records. The near-random treatment assignment (propensity AUC = 0.509) provides strong internal validity for causal estimation. Evaluated via Qini coefficient and cumulative gain curves, the S-Learner achieves the highest Qini score of 0.376, with the top 20\% of customers ranked by predicted CATE capturing 77.7\% of all incremental conversions, a 3.9$\times$ improvement over random targeting. SHAP analysis identifies f8 as the dominant heterogeneous treatment effect (HTE) driver among the 12 anonymized covariates. Causal Forest uncertainty quantification reveals that 1.9\% of customers are confident persuadables (lower 95\% CI $> 0$) and 0.1\% are confident sleeping dogs (upper 95\% CI $< 0$). Our results provide practitioners with evidence-based guidance on method selection for large-scale uplift modeling pipelines.
\end{abstract}

\section{Introduction}

Standard A/B testing yields Average Treatment Effects (ATE), campaign-level estimates that cannot distinguish between customers who respond positively, negatively, or not at all to a marketing intervention. In large-scale digital marketing, this limitation creates three economically costly targeting errors: wasted spend on always-converters, negative returns from sleeping dogs who react adversely to treatment, and missed revenue from high-responders who are under-prioritized.

Uplift modeling, grounded in the Neyman-Rubin potential outcomes framework \cite{rubin1974}, addresses this by estimating the Conditional Average Treatment Effect:
\begin{equation}
    \tau(\mathbf{x}) = \mathbb{E}[Y(1) - Y(0) \mid \mathbf{X} = \mathbf{x}]
\end{equation}
where $Y(1)$ and $Y(0)$ are potential outcomes under treatment and control, and $\mathbf{X}$ is the covariate vector. Individual-level CATE estimates enable precision targeting: deploy treatment only where $\tau(\mathbf{x}) > \text{threshold}$.

Despite a growing literature on CATE estimation methods, systematic empirical comparisons at industrial scale remain sparse. Existing benchmarks \cite{devriendt2020,kunzel2019} are largely conducted at smaller scales or on synthetic data.

We contribute \textbf{UpliftBench}: a reproducible, large-scale benchmark comparing meta-learner and Causal Forest approaches on the Criteo Uplift v2.1 dataset, with full evaluation infrastructure including Qini scoring, policy simulation, uncertainty quantification, and SHAP-based HTE attribution.

\section{Related Work}

\paragraph{Meta-learners for CATE estimation.}
\cite{kunzel2019} introduced the X-Learner, demonstrating improved CATE estimation under treatment imbalance relative to S-Learner and T-Learner approaches. The S-Learner fits a single outcome model with treatment as a feature; the T-Learner fits separate models per treatment arm. Both are reviewed in \cite{curth2021}, who analyze their bias-variance tradeoffs theoretically.

\paragraph{Causal Forests.}
\cite{wager2018} introduced Causal Forests as a nonparametric approach to CATE estimation, providing asymptotically valid confidence intervals via honest splitting. \cite{athey2019} extended this to generalized random forests. EconML \cite{battocchi2019} provides the production implementation used in this work.

\paragraph{Uplift benchmarking.}
\cite{devriendt2020} conduct a systematic comparison of uplift modeling methods across multiple datasets, finding that ensemble-based methods generally outperform simpler approaches. \cite{gubela2019} examine evaluation metrics including Qini coefficient and AUUC in the context of marketing attribution. Our work extends this literature to the 13.98M-record scale with LightGBM-based meta-learners and full uncertainty quantification.

\paragraph{Evaluation metrics.}
The Qini coefficient \cite{radcliffe2007} measures the area between the model's cumulative gain curve and the random targeting baseline, providing a scalar ranking metric for uplift models. We adopt Qini as our primary evaluation metric, supplemented by cumulative gain curves and policy simulation.

\section{Dataset}

We use the \textbf{Criteo Uplift v2.1} dataset \cite{diemert2018}, a public benchmark released by the Criteo AI Lab for uplift modeling research. Table~\ref{tab:dataset} summarizes key properties. The near-random propensity score (AUC $\approx$ 0.51) confirms near-random treatment assignment, lending the CATE estimates high internal validity. The 85/15 treatment imbalance motivates inclusion of the X-Learner, specifically designed for imbalanced settings. We model the \textit{visit} outcome throughout; conversion modeling is left for future work given its 0.29\% base rate.

\begin{table}[t]
\centering
\small
\caption{Criteo Uplift v2.1 Dataset Summary}
\label{tab:dataset}
\begin{tabular}{ll}
\toprule
\textbf{Property} & \textbf{Value} \\
\midrule
Records           & 13,979,592 \\
Features          & 12 covariates (f0--f11) \\
Treatment         & 85\% treated / 15\% control \\
Primary Outcome   & visit (4.7\% positive rate) \\
Secondary Outcome & conversion (0.29\% rate) \\
Propensity AUC    & 0.509 \\
\bottomrule
\end{tabular}
\end{table}

\section{Methods}

\subsection{Pipeline Overview}
The full pipeline proceeds as follows: (1)~\textbf{Preprocessing}: 80/20 stratified train-test split (random\_state=42); StandardScaler normalization applied to all 12 covariates. (2)~\textbf{Propensity estimation}: Logistic Regression propensity score estimated on training data; score appended as a 13th feature for meta-learner training. (3)~\textbf{CATE estimation}: Four methods trained on the training split. (4)~\textbf{Evaluation}: Qini coefficient, cumulative gain curves, and policy simulation computed on the held-out test set. (5)~\textbf{Interpretability}: SHAP values computed on T-Learner predictions to attribute HTE to individual covariates.

\subsection{CATE Estimators}
\label{sec:estimators}

All meta-learners use \textbf{LightGBM} as the base learner, a gradient-boosted decision tree framework optimized for speed and performance on large tabular datasets.

\textbf{S-Learner.} A single outcome model $\mu(\mathbf{x}, t)$ is trained with treatment indicator $t$ as an additional feature. CATE is estimated as $\hat{\tau}(\mathbf{x}) = \mu(\mathbf{x}, 1) - \mu(\mathbf{x}, 0)$. The S-Learner regularizes toward zero treatment effect, reducing variance at the potential cost of bias.

\textbf{T-Learner.} Separate outcome models $\mu_1(\mathbf{x})$ and $\mu_0(\mathbf{x})$ are trained on treated and control populations respectively. CATE is estimated as $\hat{\tau}(\mathbf{x}) = \mu_1(\mathbf{x}) - \mu_0(\mathbf{x})$. This approach captures group-specific patterns but may overfit under the 85/15 treatment imbalance.

\textbf{X-Learner} \cite{kunzel2019}\textbf{.} A two-stage cross-fitting approach that imputes individual treatment effects by applying each group's outcome model to the opposite group, then re-fitting on the imputed effects weighted by propensity scores. Designed to handle treatment imbalance.

\textbf{Causal Forest} \cite{wager2018}\textbf{.} EconML's doubly robust Causal Forest with honest splitting, providing point estimates and 95\% bootstrap confidence intervals at the individual level. Evaluated on a 10\% stratified subsample ($\approx$280K records) due to computational constraints.

\subsection{Evaluation Metrics}
\label{sec:evaluation}

\textbf{Qini Coefficient.} Primary ranking metric. Measures the area between the model's cumulative gain curve and the random targeting baseline:
\begin{equation}
    \text{Qini} = \int_0^1 \left[ G(\phi) - \phi \cdot G(1) \right] d\phi
\end{equation}
where $G(\phi)$ is the fraction of incremental conversions captured when targeting the top $\phi$ fraction of customers ranked by descending predicted CATE.

\textbf{Policy Simulation.} Translates Qini curves into actionable budget vs. incremental conversion tradeoff estimates for campaign planning.

\textbf{Uncertainty Quantification.} Causal Forest 95\% confidence intervals segment customers into confident persuadables (lower CI $> 0$), confident sleeping dogs (upper CI $< 0$), and uncertain segments.

\section{Results}

\subsection{Model Comparison}

\begin{figure*}[t]
    \centering
    \includegraphics[width=0.72\textwidth]{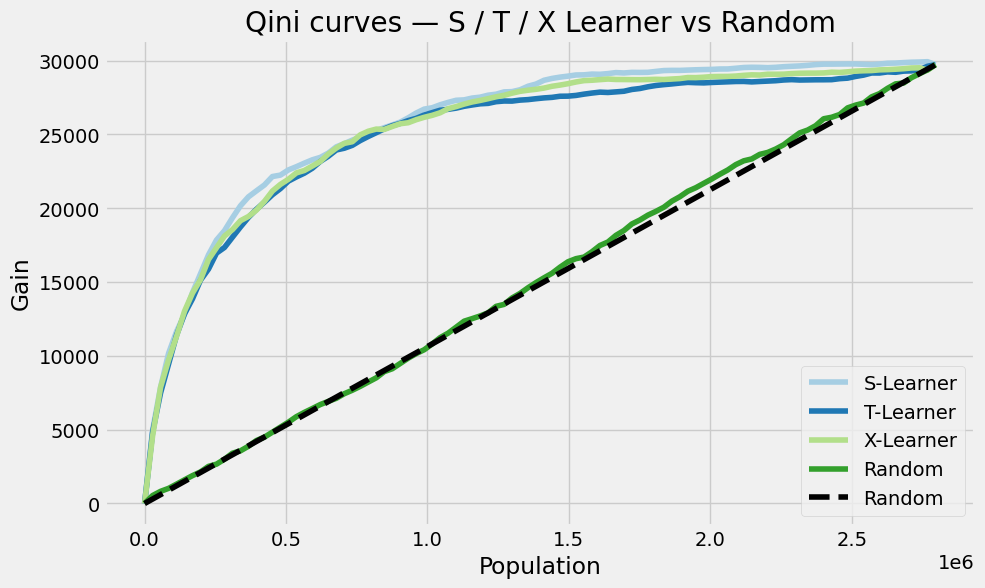}
    \caption{Qini curves for all four CATE estimators. S-Learner achieves the highest cumulative gain across all targeting thresholds.}
    \label{fig:qini}
\end{figure*}

\begin{figure*}[t]
    \centering
    \includegraphics[width=0.48\textwidth]{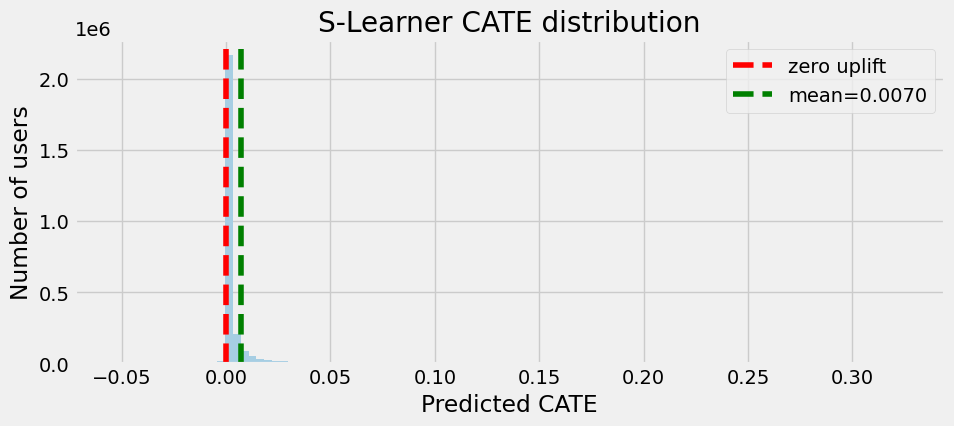}
    \hfill
    \includegraphics[width=0.48\textwidth]{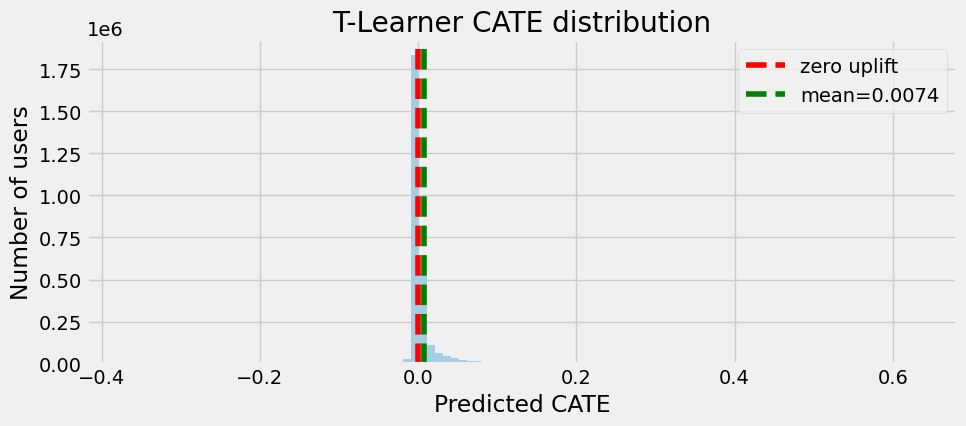}
    \caption{CATE distributions for S-Learner (left) and T-Learner (right). S-Learner exhibits significantly lower variance, consistent with its implicit regularization toward zero treatment effect.}
    \label{fig:cate_dist}
\end{figure*}

Table~\ref{tab:results} summarizes performance across all four estimators. Figure~\ref{fig:qini} shows the Qini curves; the S-Learner dominates across all targeting thresholds. Figure~\ref{fig:cate_dist} shows the CATE distributions for S-Learner and T-Learner, illustrating the variance difference.

\begin{table*}[t]
\centering
\small
\caption{Model Performance on Criteo Uplift v2.1}
\label{tab:results}
\begin{tabular}{lccccc}
\toprule
\textbf{Model} & \textbf{CATE Mean} & \textbf{CATE Std} & \textbf{Qini} & \textbf{Top 20\%} & \textbf{Top 50\%} \\
\midrule
S-Learner       & 0.0070 & 0.0223 & \textbf{0.3759} & \textbf{77.7\%} & \textbf{95.8\%} \\
X-Learner       & 0.0074 & 0.0248 & 0.3615          & 76.0\%          & 92.8\% \\
T-Learner       & 0.0074 & 0.0267 & 0.3503          & 75.1\%          & 92.4\% \\
Causal Forest   & 0.0072 & 0.0377 & 0.2524          & --              & -- \\
Random Baseline & --     & --     & $\approx$0.007  & 20.0\%          & 50.0\% \\
\bottomrule
\end{tabular}
\end{table*}

The S-Learner achieves the highest Qini score (0.3759) and top-decile capture rate with the lowest CATE variance, indicating the most stable ranking behavior for campaign targeting. Contrary to prior theoretical expectations, X-Learner does not outperform S-Learner despite the 85/15 treatment imbalance, a finding consistent with \cite{curth2021}'s analysis of finite-sample regularization effects. The Causal Forest's lower Qini score (0.2524) likely reflects subsample limitations ($\approx$280K vs. 11.2M training records) rather than a fundamental performance deficit.

\subsection{Targeting Efficiency}

\begin{figure*}[t]
    \centering
    \includegraphics[width=0.72\textwidth]{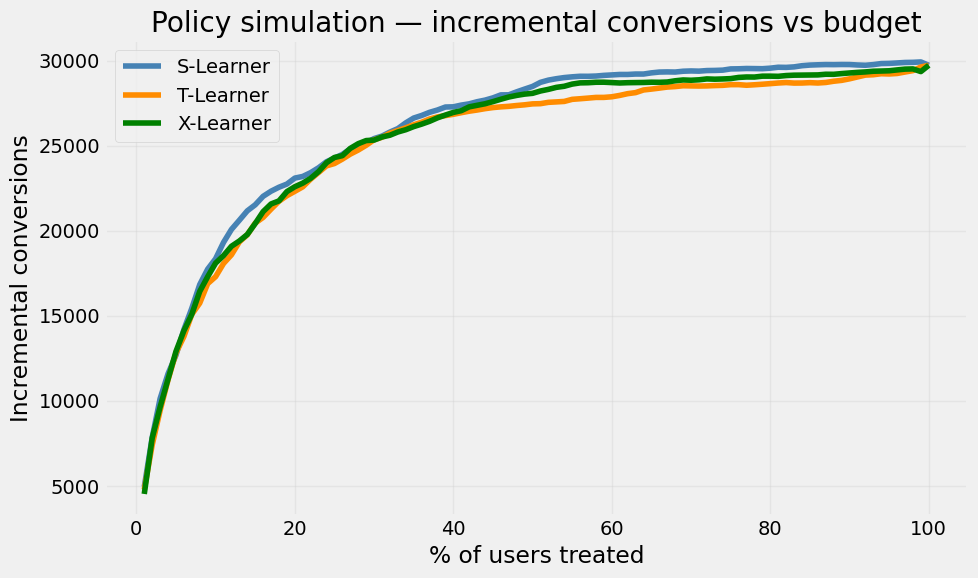}
    \caption{Policy simulation: incremental conversions captured vs. fraction of population contacted. S-Learner captures 77.7\% of all incremental conversions at a 20\% contact rate.}
    \label{fig:policy}
\end{figure*}

Table~\ref{tab:policy} and Figure~\ref{fig:policy} illustrate the business value of uplift-based targeting on a hypothetical campaign of 1,000,000 customers at \$1 per contact. Targeting the top 20\% of customers by predicted CATE achieves 3.9$\times$ the conversion efficiency of random selection at 80\% lower contact cost, directly reducing customer acquisition cost without increasing marketing spend.

\begin{table*}[t]
\centering
\small
\caption{Policy Simulation: Targeting Efficiency}
\label{tab:policy}
\begin{tabular}{lccc}
\toprule
\textbf{Strategy} & \textbf{Contacts} & \textbf{Incremental Conversions Captured} & \textbf{Efficiency vs. Random} \\
\midrule
Untargeted (100\%) & 1,000,000 & 100\%  & 1.0$\times$ \\
Top 20\% S-Learner & 200,000   & 77.7\% & \textbf{3.9$\times$} \\
Random 20\%        & 200,000   & 20.0\% & 1.0$\times$ \\
\bottomrule
\end{tabular}
\end{table*}

\subsection{Uncertainty Quantification}

\begin{figure*}[t]
    \centering
    \includegraphics[width=0.85\textwidth]{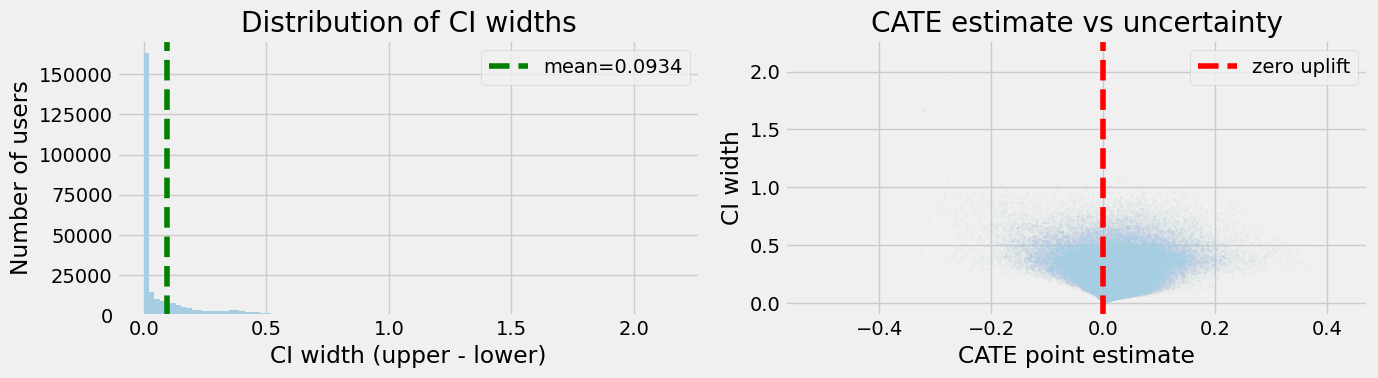}
    \caption{Causal Forest 95\% CI distribution on the $\approx$280K evaluation subsample. Left: distribution of CI widths. Right: CATE point estimates vs. uncertainty. Most customers have uncertain effect estimates; 1.9\% are confident persuadables.}
    \label{fig:cfuncert}
\end{figure*}

Causal Forest 95\% confidence intervals on the $\approx$280K evaluation subsample yield the segmentation in Table~\ref{tab:uncertainty}. Figure~\ref{fig:cfuncert} shows the distribution of CI widths and CATE point estimates. The predominance of uncertain estimates (98\%) reflects the fundamental problem of causal inference at fine granularity \cite{holland1986}. Probabilistic targeting using the full CATE distribution outperforms binary rule-based approaches derived from confidence intervals alone.

\begin{table}[t]
\centering
\small
\caption{Causal Forest Customer Segmentation}
\label{tab:uncertainty}
\begin{tabular}{lcc}
\toprule
\textbf{Segment} & \textbf{Criterion} & \textbf{Share} \\
\midrule
Confident persuadables  & Lower CI $> 0$ & 1.9\% \\
Confident sleeping dogs & Upper CI $< 0$ & 0.1\% \\
Uncertain               & CI overlaps 0  & 98.0\% \\
\bottomrule
\end{tabular}
\end{table}

\subsection{Feature Attribution via SHAP}

\begin{figure}[t]
    \centering
    \includegraphics[width=\columnwidth]{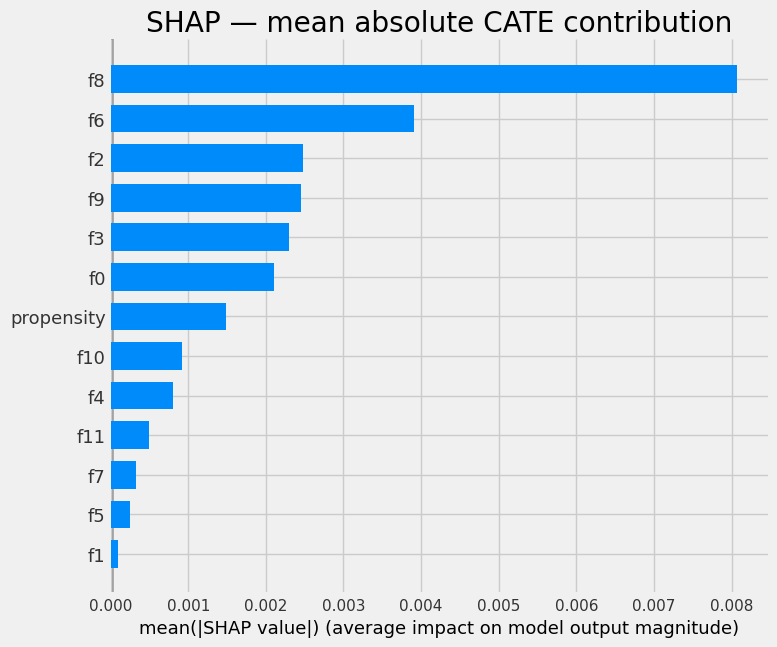}\\[4pt]
    \includegraphics[width=\columnwidth]{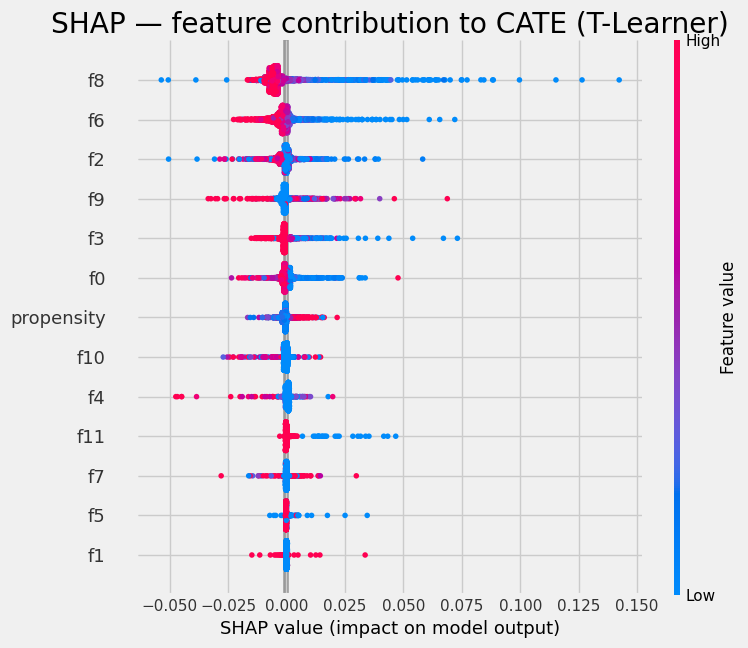}
    \caption{Top: Mean absolute SHAP values for T-Learner CATE predictions; f8, f6, and f2 dominate. Bottom: SHAP beeswarm plot; each point is one customer, color indicates feature value magnitude.}
    \label{fig:shap_bar}
    \label{fig:shap_beeswarm}
\end{figure}

SHAP analysis on T-Learner predictions identifies the relative contribution of each anonymized covariate (f0--f11) to treatment effect heterogeneity. Feature f8 is the dominant HTE driver, followed by f6 and f2, collectively accounting for the majority of CATE variance. Figure~\ref{fig:shap_bar} shows the mean absolute SHAP values (top) and the full beeswarm distribution (bottom). This attribution provides actionable signal for downstream feature selection in production targeting systems, even under feature anonymization.

\section{Discussion}

\textbf{Why does S-Learner outperform X-Learner?} The regularization induced by treating treatment as a feature in a single model may be advantageous at this scale. With 11.2M training records and a low base rate outcome (4.7\%), the implicit shrinkage of the S-Learner reduces overfitting to noisy individual-level effect estimates. Practitioners should not default to X-Learner under treatment imbalance without empirical validation at their specific data scale.

\textbf{Causal Forest at scale.} Computational constraints requiring subsampling are a practical limitation for industrial deployment. Future work should explore approximate or distributed Causal Forest implementations that scale to full-population estimation.

\textbf{Feature anonymization.} The anonymized covariates (f0--f11) limit direct business interpretation of SHAP results. Practitioners with named features would gain substantially more actionable HTE attribution.

\section{Limitations}

\begin{itemize}[noitemsep,topsep=2pt]
    \item Causal Forest evaluated on 10\% subsample; full-population estimates may differ from meta-learner results.
    \item Only the \textit{visit} outcome is modeled; conversion modeling is recommended for revenue-level impact.
    \item Feature anonymization prevents direct business interpretation of HTE drivers.
    \item The 85/15 treatment imbalance may affect T-Learner and X-Learner stability in low-density feature regions.
    \item External validity: findings are specific to the Criteo dataset and may not generalize across domains.
\end{itemize}

\section{Conclusion}

We present UpliftBench, a large-scale empirical comparison of CATE estimation methods on 13.98M customer records. Key findings: (1) S-Learner with LightGBM achieves the highest Qini score (0.376) and 77.7\% incremental conversion capture at top-20\% targeting, a 3.9$\times$ efficiency gain over random selection; (2) X-Learner does not consistently outperform S-Learner despite treatment imbalance, suggesting scale-dependent regularization effects; (3) Causal Forest provides valuable uncertainty quantification but requires subsampling at this scale; (4) SHAP analysis identifies f8 as the dominant HTE driver, enabling attribution even under feature anonymization. The full pipeline, evaluation code, and results are publicly available at \url{https://github.com/Aman12x/UpliftBench}.



\begin{thebibliography}{99}

\bibitem{athey2019}
Athey, S., Tibshirani, J., \& Wager, S. (2019).
Generalized random forests.
\textit{Annals of Statistics}, 47(2), 1148--1178.

\bibitem{battocchi2019}
Battocchi, K., et al. (2019).
EconML: A Python package for ML-based heterogeneous treatment effects estimation.
\textit{GitHub}. \url{https://github.com/microsoft/EconML}

\bibitem{curth2021}
Curth, A., \& van der Schaar, M. (2021).
Nonparametric estimation of heterogeneous treatment effects: From theory to learning algorithms.
\textit{Proceedings of AISTATS}.

\bibitem{devriendt2020}
Devriendt, F., et al. (2020).
A literature survey and experimental evaluation of the state-of-the-art in uplift modeling.
\textit{Journal of Machine Learning Research}.

\bibitem{diemert2018}
Diemert, E., Betlei, A., Renaudin, C., \& Amini, M.R. (2018).
A large scale benchmark for uplift modeling.
\textit{Proceedings of KDD}.

\bibitem{gubela2019}
Gubela, R., et al. (2019).
Conversion uplift in e-commerce.
\textit{International Journal of Information Technology \& Decision Making}.

\bibitem{holland1986}
Holland, P. W. (1986).
Statistics and causal inference.
\textit{Journal of the American Statistical Association}, 81(396), 945--960.

\bibitem{kunzel2019}
K\"{u}nzel, S. R., Sekhon, J. S., Bickel, P. J., \& Yu, B. (2019).
Metalearners for estimating heterogeneous treatment effects using machine learning.
\textit{Proceedings of the National Academy of Sciences}, 116(10), 4156--4165.

\bibitem{radcliffe2007}
Radcliffe, N. J. (2007).
Using control groups to target on predicted lift.
\textit{Direct Marketing Analytics Journal}.

\bibitem{rubin1974}
Rubin, D. B. (1974).
Estimating causal effects of treatments in randomized and nonrandomized studies.
\textit{Journal of Educational Psychology}, 66(5), 688--701.

\bibitem{wager2018}
Wager, S., \& Athey, S. (2018).
Estimation and inference of heterogeneous treatment effects using random forests.
\textit{Journal of the American Statistical Association}, 113(523), 1228--1242.

\end{thebibliography}
\end{document}